\documentstyle[12pt]{article}
\catcode`\@=11

\@addtoreset{equation}{section}
\def\theequation{\arabic{section}.\arabic{equation}}
     
\def\section{\@startsection{section}{1}{\z@}{3.5ex plus 1ex minus
   .2ex}{2.3ex plus .2ex}{\large\bf}}
   \def\thesection{\arabic{section}}

\def\appendix{\setcounter{section}{0}
        \def\thesection{Appendix\ \Alph{section}}
        \def\theequation{\Alph{section}.\arabic{equation}}}
        
\long\def\@makefntext#1{\parindent 0cm\noindent
\hbox to 1em{\hss$^{\@thefnmark}$}#1}

\def\eqnarray{\let\@currentlabel=\theequation\refstepcounter{equation}
    \global\@eqnswtrue
    \global\@eqcnt\z@\tabskip\@centering\let\\=\@eqncr
    $$\halign to \displaywidth\bgroup\@eqnsel\hskip\@centering
      $\displaystyle\tabskip\z@{##}$&\global\@eqcnt\@ne 
       \hfil${{}##{}}$\hfil
      &\global\@eqcnt\tw@ $\displaystyle\tabskip\z@{##}$\hfil 
       \tabskip\@centering&\llap{##}\tabskip\z@\cr}
\def\lefteqn#1{\hbox to 4\arraycolsep{$\displaystyle #1$\hss}}

\newcommand{\beq}{\begin{equation}}
\newcommand{\eeq}{\end{equation}}
\newcommand{\Th}{{\Theta}}
\newcommand{\om}{{\omega}}

\begin{document}

%
%
%
%
\def\citen#1{%
\edef\@tempa{\@ignspaftercomma,#1, \@end, }
\edef\@tempa{\expandafter\@ignendcommas\@tempa\@end}%
\if@filesw \immediate \write \@auxout {\string \citation {\@tempa}}\fi
\@tempcntb\m@ne \let\@h@ld\relax \let\@citea\@empty
\@for \@citeb:=\@tempa\do {\@cmpresscites}%
\@h@ld}
%
\def\@ignspaftercomma#1, {\ifx\@end#1\@empty\else
   #1,\expandafter\@ignspaftercomma\fi}
\def\@ignendcommas,#1,\@end{#1}
%
%
\def\@cmpresscites{%
 \expandafter\let \expandafter\@B@citeB \csname b@\@citeb \endcsname
 \ifx\@B@citeB\relax 
    \@h@ld\@citea\@tempcntb\m@ne{\bf ?}%
    \@warning {Citation `\@citeb ' on page \thepage \space undefined}%
 \else
    \@tempcnta\@tempcntb \advance\@tempcnta\@ne
    \setbox\z@\hbox\bgroup 
    \ifnum\z@<0\@B@citeB \relax
       \egroup \@tempcntb\@B@citeB \relax
       \else \egroup \@tempcntb\m@ne \fi
    \ifnum\@tempcnta=\@tempcntb 
       \ifx\@h@ld\relax 
          \edef \@h@ld{\@citea\@B@citeB}%
       \else 
          \edef\@h@ld{\hbox{--}\penalty\@highpenalty \@B@citeB}%
       \fi
    \else   
       \@h@ld \@citea \@B@citeB \let\@h@ld\relax
 \fi\fi%
 \let\@citea\@citepunct
}
%
\def\@citepunct{,\penalty\@highpenalty\hskip.13em plus.1em minus.1em}%
%
%
\def\@citex[#1]#2{\@cite{\citen{#2}}{#1}}%
%
%
\def\@cite#1#2{\leavevmode\unskip
  \ifnum\lastpenalty=\z@ \penalty\@highpenalty \fi 
  \ [{\multiply\@highpenalty 3 #1
      \if@tempswa,\penalty\@highpenalty\ #2\fi 
    }]\spacefactor\@m}
\let\nocitecount\relax  
%

\begin{titlepage}
\vspace{.5in}
\begin{flushright}
UCD-99-12\\
NSF-ITP-99-069\\
gr-qc/9906126\\
June 1999\\
\end{flushright}
\vspace{.5in}
\begin{center}
{\Large\bf
 Entropy from Conformal Field Theory\\[.5ex] at Killing Horizons}\\
\vspace{.4in}
{S.~C{\sc arlip}\footnote{\it email: carlip@dirac.ucdavis.edu}\\
       {\small\it Department of Physics}\\
       {\small\it University of California}\\
       {\small\it Davis, CA 95616}\\{\small\it USA}}
\end{center}

\vspace{.5in}
\begin{center}
{\large\bf Abstract}
\end{center}
\begin{center}
\begin{minipage}{4.75in}
{\small
On a manifold with boundary, the constraint algebra of general
relativity may acquire a central extension, which can be
computed using covariant phase space techniques.  When the
boundary is a (local) Killing horizon, a natural set of
boundary conditions leads to a Virasoro subalgebra with a
calculable central charge.  Conformal field theory methods
may then be used to determine the density of states at the
boundary.  I consider a number of cases---black holes, Rindler
space, de Sitter space, Taub-NUT and Taub-Bolt spaces, and 
dilaton gravity---and show that the resulting density of
states yields the expected Bekenstein-Hawking entropy.  The 
statistical mechanics of black hole entropy may thus be fixed 
by symmetry arguments, independent of details of quantum
gravity.
}
\end{minipage}
\end{center}
\end{titlepage}
\addtocounter{footnote}{-1}

In attempting to understand the thermodynamic properties of black 
holes, physicists are pulled in two directions.  On the one hand, 
we would like to find a microscopic ``statistical mechanical'' 
description of black hole thermodynamics in terms of some specific 
set of quantum states.  But we also recognize that the existing 
derivations of black hole temperature and entropy use only broad 
features of quantum field theory and semiclassical gravity, 
suggesting that a microscopic explanation cannot be too sensitive 
to the details of quantum gravity.  We seem to need a kind of 
``quantum gravity without quantum gravity'': a general principle 
that governs the density of states in quantum gravity and yet is 
independent of the details of the theory.

A natural candidate for such a general principle is a symmetry.  
The idea of using symmetry arguments to count states in quantum 
gravity was originally suggested by Strominger \cite{Strominger} in 
the context of the (2+1)-dimensional black hole.  Brown and Henneaux 
\cite{Brown1} had noted in 1986 that (2+1)-dimensional gravity with 
a negative cosmological constant has an asymptotic symmetry consisting 
of a pair of Virasoro algebras, implying that any microscopic quantum
theory should be a conformal field theory.  But conformal field 
theories have a peculiar property: the Cardy formula \cite{Cardy} 
determines the asymptotic density of states entirely in terms of 
the Virasoro algebra, independent of other details of the theory.  
Strominger observed that if one uses the central charge of ref.\ 
\cite{Brown1} in the Cardy formula, the resulting density of 
states reproduces the standard Bekenstein-Hawking entropy for the 
(2+1)-dimensional black hole, thus providing the sort of universal 
mechanism we need.

Unfortunately, there are two basic limitations to Strominger's 
approach.  First, of course, it works only in 2+1 dimensions.
This is less restrive than it might appear, since many of 
the higher dimensional black holes in string theory have
near-horizon geometries that reduce that of the (2+1)-dimensional
black hole \cite{Carlip2}.  Still, it seems unnatural to depend
on particular features of 2+1 dimensions for what ought to be a 
universal computation.  Second, since Strominger's argument is 
based on an algebra of transformations at infinity, it is 
insensitive to important details of the structure of the interior 
of spacetime, and yields only a sort of ``maximum possible entropy.''  
For example, the entropy computation of ref.\ \cite{Strominger} 
applies equally well to a black hole of mass $m$ and a spherical 
star of the same mass.  It may be that finer details of the 
conformal field theory at infinity can distinguish such cases 
\cite{Ross}, but presumably one would like to be able to count 
the states of a black hole more directly.

In ref.\ \cite{Carlip1}, I proposed a generalization of the 
Brown-Henneaux-Strominger construction based on the algebra of 
deformations at a black hole horizon.  If the horizon is treated 
as a boundary, the algebra of constraints in general relativity 
acquires a central extension.  Given a plausible set of boundary 
conditions, this extended algebra contains a natural Virasoro 
subalgebra, and the Cardy formula can again be used to obtain 
the correct entropy.  This construction is valid for black holes 
in any dimension.  Unfortunately, however, the derivation is tied 
to a particular ``Schwarzschild-like'' coordinate system, and some 
rather arbitrary restrictions on diffeomorphisms are required.  

In this paper, I rederive the central extension of the constraint
algebra of general relativity using manifestly covariant phase space 
methods.  If one takes the boundary to be a surface that looks 
locally like a Killing horizon, the resulting algebra again contains 
a natural Virasoro algebra with a calculable central charge.  I 
consider a variety of spacetimes---rotating black holes, Rindler 
space, de Sitter space, and Taub-NUT and Taub-Bolt spaces---as well 
as the extension to two-dimensional dilaton gravity.  In each case, 
the Cardy formula leads to a density of states that yields the 
expected Bekenstein-Hawking entropy. 

\section{The General Argument \label{sec0}}

Since much of this paper is rather technical, I will begin with a
summary of the general argument and a discussion of some of the
broader physical issues.  My starting point is the investigation
of the algebra of constraints in general relativity in the presence
of a boundary.  Naively, one would expect this algebra to be
equivalent to the algebra of diffeomorphisms of the spacetime
$M$.  But as Brown and Henneaux have stressed \cite{Brown1},
boundary terms in the constraints can lead to a central extension 
of $\hbox{Diff}\,M$.

Such a central extension is of interest in its own right, but it
becomes especially important if a subalgebra isomorphic to
$\hbox{Diff}\,S^1$ or $\hbox{Diff}\,{\bf R}$ acquires a central
term.  A centrally extended algebra of this type is known as a 
Virasoro algebra, and such algebras play a fundamental role in
conformal field theory.  The algebra we are considering is a
classical Poisson algebra, but any quantum theory of gravity
should presumably inherit this structure, perhaps with order $\hbar$
corrections.  This means that we can use powerful techniques 
developed in conformal field theory to obtain useful information
about quantum gravity.  In particular, the Cardy formula determines
the asymptotic behavior of the density of states in terms of
quantities fixed almost uniquely by the Virasoro algebra.  (I
discuss the derivation and applicability of the Cardy formula
in \ref{appenc}.)

To investigate the constraint algebra, I use covariant phase space
methods \cite{Ashtekar0,Crn,Barnich}, which exploit the isomorphism 
between phase space and the space of solutions of the field equations 
to provide a canonical formalism that is manifestly covariant.  In
particular, a formalism developed by Wald and his collaborators
\cite{Lee,Wald1,Wald2,Wald3} is especially well suited for my 
purposes.  The results are summarized by eqns.\ (\ref{b3}) and 
(\ref{b5}), which give the central extension of the algebra of 
constraints for an arbitrary covariant theory.

The central term in the constraint algebra arises from boundary
terms in the generators.  Since I am interested in black holes, I
specify boundary conditions that reflect the presence of a horizon. 
There are a number of ways of imposing such a requirement.  I choose
the simplest, though not the most general, which is to demand that
the boundary look locally like a Killing horizon.  This condition,
along with a somewhat more mysterious restriction on the average 
surface gravity, is sufficient to determine a central extension of 
the constraint algebra.  In particular, diffeomorphisms of the 
``$r-t$ plane,'' which play the central role in Euclidean path 
integral computation of black hole entropy, acquire a central term 
with the structure one expects for a Virasoro algebra.  The general 
form of this algebra is given by eqns.\ (\ref{c15})--(\ref{c16}); 
the specialization to a one-parameter subalgebra is given by eqn.\ 
(\ref{d9}).  If one now employs the Cardy formula, the resulting 
density of states (\ref{d13}) is precisely what is needed to reproduce 
the Bekenstein-Hawking entropy.

It is worth pausing for a moment to discuss the sense in which
one can treat a horizon as a boundary.  A black hole horizon 
is not, after all, a true ``edge'' of spacetime; there is nothing 
to stop an external observer from passing through the horizon. 
 
Consider, however, an arbitrary quantum mechanical question about 
a black hole.  Such a question automatically calls for the computation 
of a conditional probability: for instance, ``What is the probability 
of observing a photon of Hawking radiation of frequency $\nu$, 
{\em given\/} the presence of a black hole with a horizon of area
$A$?''

Now, in semiclassical gravity, such a condition can be imposed by 
fixing the background metric to be that of a prescribed black hole.  
In a true quantum theory, however, this is no longer possible,
since the metric is itself a quantum field that cannot be precisely 
specified.  Instead, the most direct way to ask such a conditional 
question is to require the presence of a surface with suitable 
properties to ensure it is a horizon.  Note that the amount of 
boundary data one is allowed to impose is ``half the phase space,''
exactly the amount compatible with the uncertainty principle.
This means that whether or not we we treat the horizon as a physical 
boundary, we must treat it as a surface upon which we impose boundary 
conditions.  The existence of such boundary conditions and the 
consequent restrictions on variations of the fields are sufficient 
to justify the methods of this paper.

The analysis I have just described was developed for black holes 
in ordinary general relativity, but its extension to other 
configurations and other theories is straightforward.  In section 
\ref{sece} I discuss some generalizations, and show that in each case 
one obtains a density of states that reproduces the expected entropy.
It thus appears that there may be a universal statistical mechanical
picture of entropy associated with horizons: regardless of the
details of a quantum theory of gravity, symmetries inherited from
the classical theory may be sufficient to determine the asymptotic
behavior of the density of states.

\section{Constraint Algebras and Covariant Phase Space \label{seca}}

Let us begin with a brief review of covariant phase space methods, in 
the formalism developed by Wald et al.\  \cite{Lee,Wald1,Wald2,Wald3}.
Consider a general diffeomorphism-invariant field theory in $n$ 
spacetime dimensions with a Lagrangian ${\bf L}[\phi]$, where 
$\bf L$ is viewed as an $n$-form and $\phi$ denotes an arbitrary 
collection of dynamical fields.  The variation of $\bf L$ takes the 
form
\beq
\delta{\bf L} = {\bf E}\cdot\delta\phi + d{\Th}
\label{a1}
\eeq
where the field equations are given by ${\bf E}=0$ and the 
symplectic potential $\Th[\phi,\delta\phi]$ is an $(n-1)$-form
determined by the ``surface terms'' in the variation of $\bf L$.
The symplectic current $(n-1)$-form $\om$ is defined by
\beq
\om[\phi,\delta_1\phi,\delta_2\phi] = 
   \delta_1\Th[\phi,\delta_2\phi] - \delta_2\Th[\phi,\delta_1\phi] ,
\label{a2}
\eeq
and its integral over a Cauchy surface $C$,
\beq
\Omega[\phi,\delta_1\phi,\delta_2\phi] =
   \int_C \om[\phi,\delta_1\phi,\delta_2\phi]
\label{a3}
\eeq
gives a presymplectic form on the space of solutions of the field
equations.  This space, in turn, can be identified with the usual 
phase space, and $\Omega$ becomes the standard presymplectic form 
of Hamiltonian mechanics \cite{Barnich,Lee}.

For any diffeomorphism generated by a smooth vector field $\xi^a$,
one can define a conserved Noether current $(n-1)$-form $\bf J$ by
\beq
{\bf J}[\xi] = \Th[\phi,{\cal L}_\xi\phi] - \xi\cdot{\bf L} ,
\label{a4}
\eeq
where ${\cal L}_\xi$ denotes the Lie derivative in the direction
$\xi$ and the dot $\cdot$ means contraction of a vector with
the first index of a form.  On shell, the Noether current is
closed, and can be written in terms of an $(n-2)$-form $\bf Q$,
the Noether charge, as
\beq
{\bf J} = d{\bf Q} .
\label{a5}
\eeq

Now consider a vector field $\xi^a$, and the corresponding generator
of diffeomorphisms $H[\xi]$.  In the covariant phase space formalism,
Hamilton's equations of motion become
\beq
\delta H[\xi] = \Omega[\phi,\delta\phi,{\cal L}_\xi\phi] .
\label{a6}
\eeq
It is easy to see that when $\phi$ satisfies the equations of motion,
\beq
\om[\phi,\delta\phi,{\cal L}_\xi\phi] 
   = \delta{\bf J}[\xi] - d(\xi\cdot\Th[\phi,\delta\phi]) ,
\label{a7}
\eeq
so by eqns.\ (\ref{a3}) and (\ref{a5}),
\beq
H[\xi] = \int_{\partial C} ({\bf Q}[\xi] - \xi\cdot{\bf B} ) ,
\label{a8}
\eeq
where the $(n-1)$-form $\bf B$ is defined by the requirement that
\beq
\delta\int_{\partial C} \xi\cdot{\bf B}[\phi] 
   = \int_{\partial C} \xi\cdot\Th[\phi,\delta\phi] .
\label{a9}
\eeq
Given a choice of boundary conditions at $\partial C$, finding 
$\bf B$ is roughly equivalent to finding the appropriate boundary 
terms for the Hamiltonian constraint in the standard ADM formalism 
of general relativity.  It should be emphasized that $\bf B$ may
not always exist; given a choice of boundary conditions, eqn.\
(\ref{a9}) may not have a solution for every vector field $\xi^a$.

For general relativity in $n$ spacetime dimensions, the Lagrangian 
$n$-form is 
\beq
{\bf L}_{a_1\dots a_n} = {1\over16\pi G}\epsilon_{a_1\dots a_n}R,
\label{a10}
\eeq
which yields a symplectic potential $(n-1)$-form \cite{Wald2}
\beq
\Th_{a_1\dots a_{n-1}}[g,\delta g] = {1\over16\pi G}
   \epsilon_{ba_1\dots a_{n-1}}\left( g^{bc}\nabla_c(g_{de}\delta g^{de}) 
   - \nabla_c\delta g^{bc}\right) .
\label{a11}
\eeq
The corresponding Noether charge, evaluated when the vacuum field 
equations hold, is
\beq
{\bf Q}_{a_1\dots a_{n-2}}[g,\xi] = -{1\over16\pi G}
   \epsilon_{bca_1\dots a_{n-2}}\nabla^b\xi^c .
\label{a12}
\eeq

\section{Central Terms in the Algebra of Diffeomorphisms \label{secb}}

In the absence of a boundary, the Poisson brackets of the generators 
$H[\xi]$ form the standard ``surface deformation algebra'' 
\cite{Teitelboim}, equivalent on shell to the algebra of diffeomorphisms.  
On a manifold with boundary, however, the addition of boundary terms 
can alter the Poisson brackets, leading to a central extension of the 
surface deformation algebra \cite{Brown1}.  That is, the Poisson algebra 
may take the form
\beq
\{H[\xi_1],H[\xi_2]\} = H[\{\xi_1,\xi_2\}] + K[\xi_1,\xi_2],
\label{b1}
\eeq
where the central term $K[\xi_1,\xi_2]$ depends on the
dynamical fields only through their (fixed) boundary values.
This phenomenon is not peculiar to gravity \cite{Arnold}; it
occurs because the generators are unique only up to the
addition of constants, and the constant term in the boundary
contribution to $H[\{\xi_1,\xi_2\}]$ may not match the
corresponding term in $\{H[\xi_1],H[\xi_2]\}$.  The existence
of such a central extension has been studied extensively in
(2+1)-dimensional gravity, both in the metric formulation
\cite{Brown1,Brown2} and in the Chern-Simons formulation
\cite{Banados1,Banados2}.  Here we wish to investigate it in 
a more general setting.

Consider the Poisson brackets of the generators of diffeomorphisms
in the covariant phase space formalism of the preceding section.  
Let $\xi_1^a$ and $\xi_2^a$ be two vector fields, and suppose the
fields $\phi$ solve the equations of motion (so, in particular, the 
``bulk'' constraints are all zero).  Denote by $\delta_{\xi}$ the 
variation corresponding to a diffeomorphism generated by $\xi$.  
For the Noether current ${\bf J}[\xi_1]$, 
\beq
\delta_{\xi_2}{\bf J}[\xi_1] = {\cal L}_{\xi_2}{\bf J}[\xi_1]
  = \xi_2\cdot d{\bf J}[\xi_1] + d(\xi_2\cdot{\bf J}[\xi_1])
  = d\left[ \xi_2\cdot(\Th[\phi,{\cal L}_{\xi_1}\phi] 
  - \xi_1\cdot{\bf L}) \right] ,
\label{b2}
\eeq
where I have used the fact that $d{\bf J} = 0$ on shell.  Hence
from eqns.\ (\ref{a6}) and (\ref{a7}),
\begin{eqnarray}
\delta_{\xi_2}H[\xi_1] &=& \int_C \delta_{\xi_2}{\bf J}[\xi_1] 
   - d(\xi_1\cdot\Th[\phi,{\cal L}_{\xi_2}\phi]) \nonumber\\
&=& \int_{\partial C} \left(\xi_2\cdot\Th[\phi,{\cal L}_{\xi_1}\phi]
   - \xi_1\cdot\Th[\phi,{\cal L}_{\xi_2}\phi] 
   - \xi_2\cdot(\xi_1\cdot{\bf L}) \right) .
\label{b3}
\end{eqnarray}

We can now take advantage of an observation due to Brown and 
Henneaux \cite{Brown1}.  Since eqn.\ (\ref{b3}) was evaluated 
on shell, the ``bulk'' part of the generator $H[\xi_1]$ on the 
left-hand side, which consists entirely of a sum of constraints, 
vanishes.  Hence the left-hand side can be interpreted as the
variation $\delta_{\xi_2}J[\xi_1]$, where $J$ is the boundary 
term in the constraint.  [I show in \ref{appenb} by direct 
computation that $\delta_{\xi_2}J[\xi_1]$ agrees with the right-hand 
side of eqn.\ (\ref{b3}).]  On the other hand, the Dirac bracket 
$\{J[\xi_1],J[\xi_2]\}^*$ means precisely the change in $J[\xi_1]$ 
under a surface deformation generated by $J[\xi_2]$; that is,
\beq
\delta_{\xi_2}J[\xi_1] = \{J[\xi_1],J[\xi_2]\}^* .
\label{b4}
\eeq
Comparing eqn.\ (\ref{b1}), evaluated on shell, we see that
\beq
K[\xi_1,\xi_2] = \delta_{\xi_2}J[\xi_1] 
   - J[\{\xi_1,\xi_2\}] ,
\label{b5}
\eeq
where $\delta_{\xi_2}J[\xi_1]$ is given by eqn.\ (\ref{b3}).
Given a suitable set of boundary conditions, this permits a simple
determination of the central term $K[\xi_1,\xi_2]$.

In particular, for vacuum general relativity, the Lagrangian $\bf L$ 
vanishes on shell, and the right-hand side of eqn.\ (\ref{b3}) can 
be computed from eqn.\ (\ref{a11}).  One obtains
\beq
\{J[\xi_1],J[\xi_2]\}^* = {1\over16\pi G}
   \int_{\partial C}\epsilon_{bca_1\dots a_{n-2}}\left[ 
   \xi_2^b\nabla_d(\nabla^d\xi_1^c - \nabla^c\xi_1^d) -
   \xi_1^b\nabla_d(\nabla^d\xi_2^c - \nabla^c\xi_2^d) \right] .
\label{b6}
\eeq

\section{Local Killing Horizons \label{secc}}

To proceed further, we must specify boundary conditions at 
$\partial C$ more precisely.  We are interested in the entropy 
associated with horizons, either black hole and cosmological, and
should thus choose boundary conditions that reflect the presence
of a horizon.  

Ashtekar et al.\ have recently discussed a general set of boundary 
conditions for isolated horizons \cite{Ashtekar}, and these, or 
their generalization to rotating horizons, may ultimately be the 
appropriate ones to use.  For now, however, I will take a more 
conservative (and easier) approach, and look for boundary conditions 
that imply the presence of a local Killing horizon.

Consider an $n$-dimensional spacetime $M$ with boundary $\partial M$,
such that a neighborhood of $\partial M$ admits a Killing vector
$\chi^a$ that satisfies $\chi^2 = g_{ab}\chi^a\chi^b = 0$ at 
$\partial M$.  $M$ need not be ``all of spacetime''---we are not 
restricting our attention to eternal stationary black holes---but 
can be a small region containing a momentarily stationary black hole 
or cosmological horizon; the condition $\chi^2=0$ can then be viewed 
as determining the location of the relevant boundary.  

In practice, it will be useful to work at a ``stretched horizon''
$\chi^2=\epsilon$, taking $\epsilon$ to zero at the end of the 
computation.  Near this stretched horizon, one can define a vector 
orthogonal to the orbits of $\chi^a$ by
\beq
\nabla_a\chi^2 = -2\kappa\rho_a ,
\label{c1}
\eeq
where $\kappa$ is the surface gravity at the horizon.  Note that
\beq
\chi^a\rho_a = -{1\over\kappa}\chi^a\chi^b\nabla_a\chi_b = 0 .
\label{c2}
\eeq
At the horizon, $\chi^a$ and $\rho^a$ become null, and the normalization
in eqn.\ (\ref{c1}) has been chosen so that $\rho^a\rightarrow\chi^a$.
Away from the horizon, however, $\chi^a$ and $\rho^a$ define two
orthogonal directions.  

If we now vary the metric, we will typically find ourselves in a
spacetime that admits no Killing vector even near $\partial M$.
In order for the boundary condition $\chi^2=0$ to continue to make
sense, we must at least require that $\chi^a\chi^b\delta g_{ab} = 0$, 
where $\chi^a$ is now viewed as a fixed vector field.  I will impose 
slightly stronger conditions, to preserve the ``asymptotic'' 
structure at the horizon:
\beq
{\chi^a\chi^b\over\chi^2}\delta g_{ab} \rightarrow 0 , \quad
\chi^a t^b\delta g_{ab} \rightarrow 0 \qquad 
\hbox{as $\chi^2\rightarrow 0$} ,
\label{c3}
\eeq
where $t^b$ is any unit spacelike vector tangent to $\partial M$.
Equivalently, we require that
\beq
\delta\chi^2 = 0 , \quad \chi^a t^b\delta g_{ab} = 0 , 
\quad\hbox{and}\ 
\delta\rho_a = -{1\over2\kappa}\nabla_a(\delta\chi^2) = 0
  \quad \hbox{at $\chi^2=0$} .
\label{c3a}
\eeq
Note that these conditions guarantee that the boundary $\chi^2=0$
remains null, and that $\chi^a$ continues to be the null normal to 
this boundary.  Indeed, the normal $\rho_a$ satisfies $\delta\rho^2
= \rho_a\rho_b\delta g^{ab}\rightarrow \chi_a\chi_b\delta g^{ab}$,
which vanishes at the boundary, and it is not hard to see that
$\delta(\rho^a-\chi^a) = \rho_b\delta g^{ab}$ has components only
along $\chi^a$ at $\partial M$.  

In viewing these boundary conditions, it may be helpful to keep a 
specific example in mind.  For a Kerr black hole in Boyer-Lindquist
coordinates, $\chi^2$ is equal to the lapse function $N^2$, and
$\chi^2 \sim h(\theta)(r-r_+)$ near the horizon $r=r_+$, where
$$
h(\theta) = {1\over m}{(m^2-a^2)^{1/2}\over m + (m^2-a^2)^{1/2}}
  (1-a\Omega_H\sin\theta) .
$$  
The first condition in eqn.\ (\ref{c3}) requires that the boundary 
remain at $r=r_+$, and that $h(\theta)$ remain fixed at the boundary
\cite{Carlip1}.  The second condition then requires that the shift
function $N^\phi$ be fixed at the boundary, or equivalently that
the angular velocity $\Omega_H$ of the horizon be held fixed.  In 
this sense, the conditions (\ref{c3}) are horizon analogs to the 
fall-off conditions one usually imposes at infinity.

For a diffeomorphism generated by a vector field $\xi^a$, eqn.\ 
(\ref{c3}) implies that
\beq
{\chi^a\chi^b\over\chi^2}\nabla_a\xi_b 
   = \chi^a\nabla_a\left({\chi_b\xi^b\over\chi^2}\right)
   - \kappa {\rho_b\xi^b\over\chi^2} = 0 .
\label{c4a}
\eeq
This suggests that we focus on vector fields of the form
\beq
\xi^a = R\rho^a + T\chi^a .
\label{c4}
\eeq
The corresponding diffeomorphisms are, in a reasonable sense, 
deformations in the ``$r$--$t$ plane,'' which are known to play a 
crucial role in the Euclidean approach to black hole thermodynamics 
\cite{Teitelboim2}.  

The appearance of a term in the ``radial'' direction $\rho^a$ may 
at first seem surprising, since we are ultimately interested in 
diffeomorphisms that preserve the horizon.  This term may be easily 
understood, though: the boundary condition $\chi^2=0$ is not quite 
diffeomorphism invariant, since $\chi^a$ is held fixed, and an 
extra transformation is necessary to restore our gauge condition 
at the boundary.  If we write the action as $I = \int{\hat\theta}
(\chi^2){\bf L}$, where $\hat\theta$ is a step function, it is not 
hard to see that, in the notation of eqn.\ (\ref{a1}),
\beq
\delta I = \int_M {\hat\theta}(\chi^2){\bf E}\cdot\delta\phi
   +\int_{\chi^2=0}\left( \Th[\phi,\delta\phi] - {1\over2\kappa}
   {\delta\chi^2\over\rho^2}\rho\cdot{\bf L} \right) ,
\label{c5a}
\eeq
where the last term comes from varying $\chi^2$ in the step
function.  The role of $R$ is essentially to remove this term,
allowing us to work with a fixed boundary even as $\chi^2$
varies.

For vector fields of the form (\ref{c4}), condition (\ref{c4a}) becomes
\beq
R = {1\over\kappa}{\chi^2\over\rho^2}\chi^a\nabla_a T .
\label{c5}
\eeq
We must now check whether the diffeomorphisms characterized by eqns.\
(\ref{c4}) and (\ref{c5}) form a closed subalgebra.  It is not hard to 
see that closure requires a new condition,
\beq
\rho^a\nabla_a T = 0
\label{c6}
\eeq
at the horizon.  

The need for this restriction can be traced back to the fact that 
eqn.\ (\ref{c5}) depends on the metric, thus making the parameters 
$\xi^a$ functions on phase space that must themselves be transformed.  
Now, it is certainly possible to work out the algebra of surface 
deformations when the $\xi^a$ are functions on phase space rather
than fixed parameters.  To do so would require adding terms in eqn.\ 
(\ref{b4}) and similar relations to reflect this additional dependence.  
For now, however, I will restrict myself to diffeomorphisms that 
satisfy condition (\ref{c6}).  For the Kerr black hole in 
Boyer-Lindquist coordinates, $\rho^a\nabla_a \sim (r-r_+)
[F_1(r,\theta)\partial_r + F_2(r,\theta)\partial_\theta]$, where 
$F_1$ and $F_2$ are well-behaved functions, so this is essentially 
a requirement that spatial derivatives not blow up at the horizon.

The boundary conditions imposed so far are fairly straightforward.
However, I show in \ref{appenb} that they are not sufficient to 
guarantee the existence of Hamiltonians $H[\xi]$ for diffeomorphisms
satisfying eqns.\ (\ref{c4}), (\ref{c5}), and (\ref{c6}).  To ensure 
integrability of eqn.\ (\ref{a9}), a further, somewhat less transparent 
condition is needed.

One possible new condition can be obtained by considering the
quantity $\tilde\kappa$ defined by
\beq
{\tilde\kappa}^2 = -{a^2\over\chi^2} ,
\label{c7a}
\eeq
where $a^a = \chi^b\nabla_b\chi^a$ is the acceleration of an orbit 
of $\chi^a$.  When $\chi^a$ is a Killing vector, it is easy to see 
that $\tilde\kappa$ approaches $\kappa$, the surface gravity, as 
$\chi^2\rightarrow0$, and that away from the horizon, $\tilde\kappa = 
\kappa\rho/|\chi|$.  Under variations of the metric, however, this will 
no longer be the case, and we cannot even demand that $\tilde\kappa$ 
be a constant.  We can, however, fix the average value of $\tilde\kappa$ 
over a cross section of the horizon, by requiring that
\beq
\delta\int_{\partial C}{\hat\epsilon}
   \left({\tilde\kappa} - {\rho\over|\chi|}\kappa\right) = 0 ,
\label{c7b}
\eeq
where $\hat\epsilon$ is the induced volume element on $\partial C$.

The technical role of this condition is discussed more fully in
\ref{appenb}, where it is shown that it guarantees the existence
of generators $H[\xi]$.  For now, let us merely note that for a
diffeomorphism of the type we are considering, condition (\ref{c7b}) 
requires that
\beq
\int_{\partial C} {\hat\epsilon}D^3 T = 0 ,
\label{c7c}
\eeq
where $D=\chi^a\nabla_a$.  For a one-parameter group of diffeomorphisms
such that $DT_\alpha = \lambda_\alpha T_\alpha$, this in turn implies 
an orthogonality relation
\beq
\int_{\partial C} {\hat\epsilon}\,
   T_\alpha T_\beta \sim \delta_{\alpha+\beta} ,
\label{c7d}
\eeq
which will be important later in our derivation of the central charge.

Now, given any one-parameter group of diffeomorphisms satisfying 
conditions (\ref{c4}), (\ref{c5}), and (\ref{c6}), with or without 
(\ref{c7c}), it is easy to check that
\beq
\{ \xi_1,\xi_2  \}^a = (T_1DT_2 - T_2DT_1)\chi^a +
  {1\over\kappa}{\chi^2\over\rho^2}D(T_1DT_2 - T_2DT_1)\rho^a .
\label{c8}
\eeq
This is isomorphic to the standard algebra of diffeomorphisms of the 
circle or the real line.  The question before us is whether the algebra 
of constraints merely reproduces this $\hbox{\it Diff\,}S^1$ or
$\hbox{\it Diff\,}{\bf R}$ algebra, or whether it acquires a central 
extension.

To compute the possible central term in the this algebra, we
return to eqns.\ (\ref{b5}) and (\ref{b6}).  Let us first consider 
the integration measure in (\ref{b6}).  Let $\cal H$ denote the 
$(n-2)$-dimensional intersection of the Cauchy surface $C$ with
the Killing horizon $\chi^2=0$.  The vector $\chi^a$ is, of course, 
one of the null normals to $\cal H$; denote the other future-directed 
null normal by $N^a$, with a normalization $N_a\chi^a=-1$.  Then 
\beq
\epsilon_{bca_1\dots a_{n-2}} = {\hat\epsilon}_{a_1\dots a_{n-2}}
(\chi_bN_c-\chi_cN_b) + \dots ,
\label{c9}
\eeq
where $\hat\epsilon$ is the induced measure on $\cal H$ and the 
omitted terms do not contribute to the integral.  In general,
we do not know much about $N^a$.  However, consider the vector
\beq
k^a = -{1\over\chi^2}\left( \chi^a - {|\chi|\over\rho}\rho^a\right) .
\label{c10}
\eeq
This vector is defined even in the limit $\chi^2\rightarrow0$; it 
is null everywhere, and is normalized so that $k_a\chi^a=-1$.  It 
follows that $N^a = k^a -\alpha\chi^a - t^a$, where $t^a$ is tangent 
to $\cal H$ and has a norm $t^2=2\alpha-\alpha^2\chi^2$.  It is then 
easy to see that
\begin{eqnarray}
\chi^b(\chi_bN_c-\chi_cN_b) &=& {|\chi|\over\rho}\rho_c - \chi^2t_c
   \nonumber\\
\rho^b(\chi_bN_c-\chi_cN_b) &=& 
   \left({\rho\over|\chi|} + t\cdot\rho\right)\chi_c .
\label{c11}
\end{eqnarray}
Thus for a vector of the form (\ref{c4}),
\beq
\xi^b\epsilon_{bca_1\dots a_{n-2}} = {\hat\epsilon}_{a_1\dots a_{n-2}}
   \left[ {|\chi|\over\rho}T\rho_c 
   + \left({\rho\over|\chi|} + t\cdot\rho\right)R\chi_c \right] 
   + O(\chi^2) .
\label{c12}
\eeq

The computation of the remainder of the integrand in eqn.\ (\ref{b6}) 
is straightforward.  It turns out that the term proportional to
$R$ in eqn.\ (\ref{c12}) gives a contribution of order $\chi^2$,
so the vector $t$ drops out of the result.  Using the identities in 
\ref{appena}, one finds that
\beq
\{J[\xi_1],J[\xi_2]\}^* = -{1\over16\pi G}\int_{\cal H}
   {\hat\epsilon}_{a_1\dots a_{n-2}}\left[
   {1\over\kappa}(T_1D^3T_2 - T_2D^3T_1) - 2\kappa(T_1DT_2-T_2DT_1)
   \right] ,
\label{c13}
\eeq
where terms of order $\chi^2$ have been omitted.

This expression has the characteristic three-derivative structure of 
the central term of a Virasoro algebra.  According to eqn.\ (\ref{b5}), 
though, we must also compute the surface term $J[\{\xi_1,\xi_2\}]$ of 
the Hamiltonian to obtain the complete expression for the central term
in the constraint algebra.  From eqn.\ (\ref{a8}), this Hamiltonian 
consists of two terms.  The first is straightforward to compute: using 
the same methods that led to eqn.\ (\ref{c12}), one finds that
\beq
{\bf Q}_{a_1\dots a_{n-2}} = {1\over16\pi G}
   {\hat\epsilon}_{a_1\dots a_{n-2}}\left( 
   2\kappa T - {1\over\kappa}D^2T \right) + O(\chi^2) .
\label{c14}
\eeq
The second term is more complicated, and is discussed in detail in
\ref{appenb}, where it is shown that it makes no further contribution.  
Hence combining eqns.\ (\ref{c13}) and (\ref{c14}), we obtain a central 
term
\beq
K[\xi_1,\xi_2] =  {1\over16\pi G}\int_{\cal H}
   {\hat\epsilon}_{a_1\dots a_{n-2}}{1\over\kappa} \left(
   DT_1D^2T_2 - DT_2D^2T_1 \right) ,
\label{c15}
\eeq
and a centrally extended constraint algebra
\beq
\{ J[\xi_1], J[\xi_2] \}^* = J[\{\xi_1,\xi_2\}]
   + K[\xi_1,\xi_2] .
\label{c16}
\eeq

\section{Counting States \label{secd}}

Equations (\ref{c8}) and (\ref{c15})--(\ref{c16}) are {\em almost\/} 
the standard Virasoro algebra for diffeomorphisms of the circle or
the real line.  This algebra consists of vectors $\xi(z)$ and generators 
$L[\xi]$ with Poisson brackets
\beq
i\{ L[\xi_1], L[\xi_2] \} = L[\{\xi_1,\xi_2\}] + {c\over24}\int 
{dz\over2\pi i}(\xi_1'\xi_2^{\prime\prime} - \xi_1'\xi_2^{\prime\prime}) 
\label{d1}
\eeq
for a constant $c$, the central charge.  The only essential difference 
between (\ref{c15})--(\ref{c16}) and (\ref{d1}) is the form of the 
integral on the right-hand side of eqn.\ (\ref{c15}).  If we let $v$ 
denote a parameter along the orbits of the Killing vector $\chi^a$, 
normalized so that $\chi^a\nabla_a v = 1$, and consider $T_1$ and $T_2$ 
to be functions of $v$ and of ``angular'' coordinates $\theta^i$ on
$\cal H$, we must require that
\beq
\int_{\cal H}{\hat\epsilon}\, T_1(v,\theta^i)T_2(v,\theta^i) 
  = \hbox{const.\,} \int dv\, T_1(v,\theta^i)T_2(v,\theta^i) 
\label{d2}
\eeq
to recover the algebra (\ref{d1}).  

Note that the left-hand side of this expression involves integration 
only over the cross section $\cal H$, and not along the orbits of 
$\chi^a$.  This mismatch of integrations was first noticed by Cadoni 
and Mignemi in the context of boundary algebras in two-dimensional 
gravity \cite{Cadoni}.  They proposed defining new generators, 
which in the notation of this paper are essentially integrals  
$\int\!dvJ$, which then form a standard Virasoro algebra.  In 
the present context, though, the meaning of such an additional $v$ 
integration is not clear.

In the absence of such an additional integration, we must choose
an ``angular'' dependence of the functions $T_i$---that is, a 
dependence on coordinates of $\cal H$---to enforce eqn.\ (\ref{d2}).  
This is precisely what the orthogonality condition (\ref{c7d}) does 
for us.  If, for example, we consider functions of $v$ with period 
$2\pi/\kappa$, as suggested by the Euclidean theory, and write our 
modes as
\beq
T_n(v,\theta^i) = {1\over\kappa}e^{in\kappa v}f_n(\theta^i) ,
\label{d3}
\eeq
then (\ref{c7d}) requires that
\beq
\int_{\cal H} {\hat\epsilon} f_m f_n \sim \delta_{m+n} ,
\label{d4}
\eeq
which in turn reproduces eqn.\ (\ref{d2}). 

In particular, for a rotating stationary black hole, the Killing 
vector $\chi^a$ that becomes null at the horizon is
\beq
\chi^a = t^a + \sum\Omega_{(\alpha)}\psi_{(\alpha)}^a , 
\label{d5}
\eeq
where $t^a$ is the Killing vector corresponding to time translation
invariance, $\psi_{(\alpha)}^a$ are the Killing vectors for rotational 
symmetry\footnote{In four spacetime dimensions, there is only
one $\psi^a$, but in higher dimensions, rotations in orthogonal
planes can commute, and distinct axial symmetries are allowed
\cite{Meyers}.} with corresponding angles $\phi_{(\alpha)}$, and 
the $\Omega_{(\alpha)}$ are angular velocities of the horizon.  A 
one-parameter group of diffeomorphisms satisfying (\ref{d4}) that 
closes under the brackets (\ref{c8}) is then given by
\beq
T_n = {1\over\kappa}\exp\left\{ in\left( \kappa v + \sum_\alpha
   \ell_{\alpha}(\phi_{(\alpha)} - \Omega_{(\alpha)}v) \right)\right\} ,
\label{d6}
\eeq
where the $\ell_{\alpha}$ are arbitrary integers, at least one of 
which must be nonzero, and the normalization has been chosen so that
\beq
\{ T_m, T_n \} = -i(m-n)T_{m+n}
\label{d7}
\eeq
in the brackets (\ref{c8}).

Diffeomorphisms of this form were first considered in ref.\ 
\cite{Carlip1}, where the angular dependence was introduced as an
ad hoc requirement.  At first sight, the specialization to this 
particular subgroup of diffeomorphisms seems artificial, but it is 
shown in \ref{appenb} that such a restriction---or more properly, 
the orthogonality relation (\ref{d4})---is forced upon us by the 
requirement that the generator $H[\xi]$ be well defined.  This is
perhaps not too surprising: the conventional Virasoro algebra is
essentially the only central extension of $\hbox{\it Diff}\,S^1$,
so one should expect the requirement of consistency to lead to such
an algebra.  We saw in section \ref{secc} that a suitable orthogonality 
condition can arise naturally from a boundary condition like (\ref{c7b}) 
that restricts the horizon integral of the surface gravity.   But
the origin of such a boundary condition remains unclear.  I will 
return to this issue in the conclusion.

Assuming the othogonality relation (\ref{d4}), it is easy to see that 
the algebra (\ref{c16}) is now a conventional Virasoro algebra.  The 
microscopic degrees of freedom, whatever their detailed characteristics, 
must transform under a representation of this algebra.  But as Strominger
observed \cite{Strominger}, this means that these degrees of freedom 
have a conformal field theoretic description, and powerful methods 
from conformal field theory are available to analyze their properties.

In particular, we can now use the Cardy formula to count states.  
If we consider modes of the form (\ref{d6}), the central term (\ref{c15}) 
is easily evaluated:
\beq
K[T_m,T_n] = -{iA\over8\pi G}m^3\delta_{m+n,0} ,
\label{d8}
\eeq
where $A$ is the area of the cross section $\cal H$.  The algebra
(\ref{c16}) thus becomes
\beq
i\{J[T_m],J[T_n]\}^* 
  = (m-n)J[T_{m+n}] + {A\over8\pi G}m^3\delta_{m+n,0} ,
\label{d9}
\eeq
which is the standard form for a Virasoro algebra with central
charge
\beq
{c\over12} = {A\over8\pi G} .
\label{d10}
\eeq
The Cardy formula also requires that we know the value of the boundary 
term $J[T_0]$ of the Hamiltonian.  This can be computed from eqn.\ 
(\ref{c14}):
\beq
J[T_0] = {A\over8\pi G} ,
\label{d11}
\eeq
where I have used the results of \ref{appenb} to justify neglecting 
the second term in eqn.\ (\ref{a8}).  

The Cardy formula then tells us that for any conformal field 
theory that provides a representation of the Virasoro algebra 
(\ref{d9})---modulo certain assumptions discussed in \ref{appenc}---the 
number of states with a given eigenvalue $\Delta$ of $J[T_0]$ grows 
asymptotically for large $\Delta$ as
\beq
\rho(\Delta) \sim 
\exp\left\{ 2\pi \sqrt{{c\over6}\left(\Delta-{c\over24}\right)} \right\} .
\label{d12}
\eeq
Inserting eqns.\ (\ref{d10}) and (\ref{d11}), we find that
\beq
\log\rho \sim {A\over4G} ,
\label{d13}
\eeq
giving the expected behavior of the entropy of a black hole.

\section{Some Examples \label{sece}}

The derivation in the preceding section focused on the entropy of 
stationary black holes in ordinary general relativity.  But it is 
easily generalized to a number of other interesting configurations.
In this section I briefly discuss some of these.

\subsection{Rindler Space \label{secea}}

A uniformly accelerated observer perceives a Killing horizon that is 
locally identical to that of a black hole.  Since the derivation above
required only local information about the horizon, it applies equally
well to Rindler space.  Subject to appropriate boundary conditions,
quantum gravitational states in Rindler space must transform under a
representation of the Virasoro algebra (\ref{d9}), and the density of
states should again be governed by (\ref{d13}), which should now be
interpreted as giving an entropy per unit horizon area.

Whether this is a reasonable result is a matter of debate in the
literature.  It seems inevitable, however, that any local description
of black hole entropy in terms of horizon observables will apply
to Rindler space as well.  The advantage of the present approach 
is that entropy is defined relative to a boundary.  For Rindler
space, the degrees of freedom counted by eqn.\ (\ref{d13}) are relevant 
only if one imposes suitable boundary conditions at the horizon.  Since 
these boundary conditions imply that information really is ``lost'' 
when it passes through the horizon, it is perhaps not unreasonable to 
attribute an entropy to the horizon.

This example illustrates a somewhat counterintuitive feature of 
quantum theory: the existence of a boundary can sometimes increase
the number of degrees of freedom.  For topological quantum field 
theories, this phenomenon has been studied in detail \cite{Barrett}.  
In Chern-Simons theories, for example, a ``bulk'' theory with only 
finitely many degrees of freedom can induce a Wess-Zumino-Witten 
model with infinitely many degrees of freedom on a boundary \cite{EMSS}. 
For these theories, the origin of the new degrees of freedom is 
understood: because boundary conditions limit admissible gauge 
transformations, quantities that would be considered ``pure gauge'' 
in the bulk become independent physical degrees of freedom on the 
boundary \cite{Carlip4,Carlip5}.  It is also possible to trace what 
happens when a boundary is eliminated, for example by ``gluing'' 
fields on two sides of a surface and summing over boundary values 
\cite{Witten}.  In that event, the full gauge invariance is restored, 
and the added symmetries lead to a reduction in the number of physical
degrees of freedom.  

While a full analysis of this sort is not yet available for quantum 
gravity, these examples suggest a way to make sense of the idea 
that Rindler space has a higher entropy than flat Minkowski space.
Rindler space is equivalent to a wedge of Minkowski space, but with
additional boundary conditions that are not present in the full
Minkowski space.  The resulting boundary degrees of freedom
presumably disappear when one glues back the rest of Minkowski
space and sums over boundary values, thereby eliminating the effect
of the boundary conditions.

\subsection{de Sitter Space \label{seceb}}

The methods introduced here may also be applied to cosmological 
horizons in de Sitter space \cite{Lin}.  The de Sitter metric in 
stationary coordinates can be written as
\beq
ds^2 = -\left(1-{r^2\over\ell^2}\right)dt^2 + 
   \left(1-{r^2\over\ell^2}\right)^{-1}dr^2 + r^2d\Omega^2 ,
\label{eb1}
\eeq
where $\Lambda=3/\ell^2$ in four space time dimensions.  The horizon 
at $r=\ell$ is a Killing horizon for the Killing vector
\beq
\chi^a = \left({\partial\ \over\partial t}\right)^a ,
\label{eb2}
\eeq
and the analysis of the preceding sections goes through with virtually
no changes, yielding an entropy 
\beq
S = {A_{\hbox{\scriptsize hor}}\over4G} = {3\pi\over G\Lambda} .
\label{eb3}
\eeq
Note that here, as in Rindler space, the horizon is associated with 
a particular set of observers.  For de Sitter space, however, the
existence of an associated entropy seems to be less debated; in
particular, standard Euclidean path integral methods \cite{Gibbons}
yield an entropy that agrees with that of eqn.\ (\ref{eb3}).

\subsection{Taub-NUT and Taub-Bolt Spaces \label{secec}}

Hawking and Hunter have recently investigated the entropies of 
Taub-NUT and Taub-Bolt spaces in the Euclidean path integral approach
\cite{Hawking1,Hawking2}.  When analytically continued to Riemannian 
signature, these spaces have metrics of the form
\beq
ds^2 = V\left(dt + 4n\cos^2{\theta\over2}d\phi\right)^2
     + V^{-1}dr^2 + (r^2-n^2)(d\theta^2 + \sin^2\theta\,d\phi^2) ,
\label{ec1}
\eeq
where $V$ is a function of $r$ and $n$ is a constant, the NUT
charge.  The metric has a string singularity along the positive $z$
axis (i.e., at $\theta=0$), a ``Misner string,'' whose existence is
signaled by the fact that a small loop around the axis does not shrink 
to zero proper length.  The Killing vector
\beq
\chi^a = \left({\partial\ \over\partial t}\right)^a -
   {1\over4n}\left({\partial\ \over\partial\phi}\right)^a
\label{ec2}
\eeq
has a norm that vanishes at $\theta=0$, and its Killing horizon 
consequently has a one-dimensional component along the positive 
$z$ axis.

One can now define a stretched horizon around the Misner string at 
$\chi^2=\epsilon$ and proceed exactly as above.  The surface gravity
$\kappa$ in eqn.\ (\ref{c1}) may be fixed by requiring that $\rho^2
+ \chi^2 \rightarrow0$ at the horizon; the result is that
\beq
\kappa = {1\over4n} ,
\label{ec3}
\eeq
yielding the correct $8\pi n$ periodicity for the Euclidean theory.
The remainder of the derivation is essentially unchanged.  (Details
will be published elsewhere \cite{Carlip3}.)  Using the Cardy formula
to count states on the string, one obtains a formal expression
\beq
S^{\hbox{\scriptsize string}} 
   = {A^{\hbox{\scriptsize string}}\over4G} ,
\label{ec4}
\eeq
for the entropy, where the induced volume element at $\theta=0$, 
$t=\hbox{const.}$ is
\beq
\hat\epsilon = 4ndrd\phi ,
\label{ec5}
\eeq
and correspondingly
\beq
A^{\hbox{\scriptsize string}} = 8\pi n \int_{r_0}^\infty dr .
\label{ec6}
\eeq
An additional contribution to $S$ comes from the ``bolt,'' which
is a horizon for the Killing vector $\tilde\chi = \partial/\partial t$.
($\chi$ and $\tilde\chi$ have in common that the lapse function
vanishes at their horizons.)

As in ref.\ \cite{Hawking1}, expression (\ref{ec6})  is divergent.  
Again, however, as in \cite{Hawking1}, one can compare the entropy 
of Taub-Bolt space to that of a reference Taub-NUT space.  Combining 
contributions from the Misner string and the ``bolt,'' one finds
\beq
S_{\hbox{\scriptsize Taub-Bolt}} - S_{\hbox{\scriptsize Taub-NUT}}
  = {1\over4G}\left[ A^{\hbox{\scriptsize bolt}} 
  + A^{\hbox{\scriptsize string}}_{\hbox{\scriptsize Taub-Bolt}}
  - A^{\hbox{\scriptsize string}}_{\hbox{\scriptsize Taub-NUT}}
  \right] = {\pi n^2\over G} ,
\label{ec7}
\eeq
in agreement with the results of Hawking and Hunter.

It should be possible to extend these results to the asymptotically
anti-de Sitter case discussed in ref.\ \cite{Hawking2}.  There have 
also been several recent attempts to regulate the Taub-NUT and
Taub-Bolt entropy by adding counterterms at infinity \cite{Mann,%
Emparan}; it would be interesting to understand these in the light
of the conformal field theory methods described here.  Work on
these issues is in progress.

\subsection{Dilaton Gravity \label{seced}}

So far, we have only looked at standard general relativity.  But
the methods of this paper can be easily extended to other covariant
theories of gravity.  As an example, let us consider a general
two-dimensional dilaton gravity theory, as described by Gegenberg,
Kunstatter, and Louis-Martinez \cite{Kunstatter}. 

After suitable field redefinitions, the Lagrangian two-form for this
model takes the form
\beq
{\bf L}_{ab} = {1\over2G}\epsilon_{ab}\left( \phi R +
   {1\over\ell^2}V(\phi) \right) ,
\label{ed1}
\eeq
where $V$ is an arbitrary function of the dilaton field $\phi$.
(The kinetic term for $\phi$ has been absorbed into $\phi R$ by a Weyl
rescaling of the metric.)  Black hole solutions are characterized by
a Killing vector
\beq
\chi^a = {\ell\over\sqrt{-g}}\epsilon^{ab}\nabla_b\phi 
\label{ed2}
\eeq
whose norm vanishes at the horizon $\phi = \phi_0$, i.e,,
\beq
\chi^2(\phi_0) = -\ell^2 g^{ab}\nabla_a\phi\nabla_b\phi\left|_{\phi=\phi_0}
   \right.  = 0 .  
\label{ed3}
\eeq
The vector $\rho^a$ is fixed near the horizon by the orthogonality
condition $\rho_a\chi^a = 0$ and the requirement that $\rho^2/\chi^2
\rightarrow -1$; it is
\beq
\rho_a = \ell\nabla_a\phi + O(\chi^2) .
\label{ed4}
\eeq
This can be checked explicitly from the solutions in ref.\ 
\cite{Kunstatter}.  Note that
\begin{eqnarray}
\chi^a\nabla_a\phi = 0 \qquad && \hbox{everywhere} \nonumber\\
\rho^a\nabla_a\phi\rightarrow0 \qquad && \hbox{at the horizon} ,
\label{ed5}
\end{eqnarray}
where the second line follows from eqn.\ (\ref{ed3}).

The symplectic potential $\Th_a$ is easily determined from the 
definition (\ref{a1}).  One obtains
\beq
\Th_a = 8\pi\phi\Th_a^{\hbox{\scriptsize grav}} 
   + {1\over2G}\epsilon_{ab}\left[ \nabla^a\phi\, g_{bc}\delta g^{bc} 
   - \nabla_c\phi\,\delta g^{bc} \right] ,
\label{ed6}
\eeq
where $\Th^{\hbox{\scriptsize grav}}$ is the symplectic potential 
(\ref{a11}) for Einstein gravity in two dimensions.  Similarly,
the Noether charge takes the form
\beq
{\bf Q}[\xi] = 8\pi\phi{\bf Q}^{\hbox{\scriptsize grav}} [\xi]
   + {1\over G}\xi^c\epsilon_{bc}\nabla^b\phi ,
\label{ed7}
\eeq
where ${\bf Q}^{\hbox{\scriptsize grav}}$ is given by eqn.\
(\ref{a12}).  

We can now use eqns.\ (\ref{b3})--(\ref{b5}) to evaluate the central
term in the constraint algebra at the horizon.  From eqn.\ (\ref{ed5}),
we see that the second term in (\ref{ed6}) gives no contribution to
$K[\xi_1,\xi_2]$.  Similarly, the second term in eqn.\ (\ref{ed7}) 
vanishes at the horizon.  We thus find that
\begin{eqnarray}
K[\xi_1,\xi_2] &=& 8\pi\phi_0 K^{\hbox{\scriptsize grav}}[\xi_1,\xi_2]
   \nonumber\\
J[\xi_0] &=& 8\pi\phi_0 J^{\hbox{\scriptsize grav}}[\xi_0] ,
\label{ed8}
\end{eqnarray}
where $K^{\hbox{\scriptsize grav}}$ is given by eqn.\ (\ref{c15}).

We must now confront the ``othogonality problem'' discussed at the
beginning of section \ref{secd}.  The boundary $\cal H$ is now a
point, so there are no ``angular'' integrals with which to impose
condition (\ref{d2}).  This is precisely the problem faced by
Cadoni and Mignemi \cite{Cadoni} at the boundary of two-dimensional
asymptotically anti-de Sitter space, and it presumably reflects
some of the difficulties in applying the AdS/CFT correspondence in
two dimensions \cite{Strominger2}.  We can proceed as in ref.\
\cite{Cadoni}, by either defining new integrated generators
$\int dv\,J[\xi]$ or by interpreting the Lagrangian (\ref{ed1})
as one coming from dimensional reduction, with hidden ``angular'' 
dependence.  

With either choice, it is straightforward to repeat the analysis of 
section \ref{secd}.  Thanks to the relation (\ref{ed8}), this last 
step is trivial; we can simply substitute (\ref{ed8}) into our previous 
results, to find
\beq
\log\rho \sim {8\pi\phi_0\over4G} ,
\label{ed9}
\eeq
which is precisely the entropy obtained by Gegenberg et al.\ 
\cite{Kunstatter}. 

\section{Conclusions and Open Questions \label{secf}}

This paper began with a puzzle: how can the microscopic states of 
responsible for black hole thermodynamics ``know'' about the results 
of semiclassical computations temperature and entropy?  I have
suggested a possible answer: the symmetries of classical general 
relativity may be powerful enough to determine the asymptotic
behavior of the density of states in any quantum theory of gravity,
independent of the microscopic details.  This is perhaps an unusual
role for a group of symmetries, but it is not unheard of; indeed,
in two-dimensional conformal field theory it is commonplace to use
the Virasoro algebra and the Cardy formula to count states.

Clearly, the most serious technical shortcoming of this analysis 
is the poor understanding of the orthogonality conditions (\ref{c7d}) 
and (\ref{d4}), which are necessary for the existence of a canonical 
Hamiltonian and a Virasoro algebra.  We saw that a boundary condition 
like that of eqn.\ (\ref{c7b}), which fixes an integral over the horizon, 
can lead to such orthogonality relations, but the argument is rather 
indirect, and seems to break down for two-dimensional theories.  It
seems likely that conditions (\ref{c7d}) and (\ref{d4}) have a deeper 
significance that is not yet understood.  In string theory, similar 
relations arise because black holes are often really compactified 
black strings; the integration that leads to the orthogonality in 
(\ref{d4}) is an integration over a compact dimension.  In standard
general relativity, it would be interesting to investigate the algebra 
of constraints in a null surface formulation \cite{null,null2}, 
in which integrals along the horizon like those appearing in eqn.\ 
(\ref{d2}) might arise more naturally.  Unfortunately, such an 
extension is not easy, since the constraint algebra on a null 
surface involves second class constraints.

Several obvious generalizations of this work should be possible.
First, the boundary condition I have chosen---the existence of a
local Killing horizon---is by no means the most general; it would
be interesting to understand the application of these techniques
to, for example, Ashtekar's ``isolated horizons'' \cite{Ashtekar}.
It should also be straightforward to extend these methods to a 
much wider class of gravitational theories, perhaps obtaining the 
generalized entropy formula of ref.\ \cite{Wald2}.

Finally, a crucial step would be to extend the methods developed 
here to dynamical black holes.  By choosing as my boundary the 
Killing horizon for a fixed Killing vector, I have implicitly ruled 
out dynamical processes such as black hole evaporation that require 
an evolving horizon.  Strominger's approach \cite{Strominger}, by 
way of contrast, leads to a single Virasoro algebra that incorporates
states corresponding to black holes with many masses and spins, but 
it does so by imposing boundary conditions at infinity rather than 
at the horizon.  Ideally, one would like to combine these two 
approaches, finding boundary conditions that refer to a particular 
horizon---thus isolating the degrees of freedom of a specific black 
hole---but that are also loose enough to allow that black hole to 
evolve in time.

\vspace{1.5ex}
\begin{flushleft}
\large\bf Acknowledgements
\end{flushleft}

Much of this research was performed at the Institute for Theoretical
Physics, Santa Barbara.  It was supported in part by the National
Science Foundation under Grant No.\ PHY94-07194, and by the Department 
of Energy under Grant No.\ DE-FG03-91ER40674.  I would like to thank 
Ted Jacobson and Bob Wald for making useful suggestions and asking 
hard questions, and Peter Beach Carlip for providing the requisite
lack of sleep.

\appendix

\section{Some Useful Identities \label{appena}}

In this appendix, I collect some useful identities involving
$\chi^a$ and $\rho^a$.  First,
\beq
\nabla_a\rho_b = -{1\over2\kappa}\nabla_a\nabla_b\chi^2 =
\nabla_b\rho_a .
\label{A1}
\eeq
\beq
\rho^a\nabla_a\chi^b - \chi^a\nabla_a\rho^b = 
-\rho^a\nabla^b\chi_a - \chi^a\nabla_a\rho^b =
\chi_a(\nabla^b\rho^a - \nabla^a\rho^b) = 0 .
\label{A2}
\eeq
\beq
\chi^a\nabla_a\chi^b = - \chi^a\nabla^b\chi_a = \kappa\rho^b .
\label{A3}
\eeq
\beq
{\chi^a\chi^b\over\chi^2}\nabla_a\rho_b 
= - {\chi^a\rho^b\over\chi^2}\nabla_a\chi_b 
= -\kappa{\rho^2\over\chi^2} .
\label{A4}
\eeq
Next, let
\beq
\chi_{[a}\nabla_b\chi_{c]} = \omega_{abc} .
\label{A5}
\eeq
Then
\beq
\omega_{abc}\omega^{abc} 
  = {1\over3}\chi^2(\nabla_a\chi_b)(\nabla^a\chi^b) 
  - {2\over3}\kappa^2\rho^2 ,
\label{A6}
\eeq
so
\begin{eqnarray}
\nabla_a\rho^a &=& {1\over\kappa}\nabla_a(\chi^b\nabla_b\chi^a)
  \nonumber\\
  &=& -{1\over\kappa}(\nabla_a\chi_b)(\nabla^a\chi^b) 
  + {1\over\kappa}R_{ab}\chi^a\chi^b 
  = -2\kappa{\rho^2\over\chi^2} + O(\chi^2)
\label{A7}
\end{eqnarray}
where the last equality uses the fact that $\omega^2/\chi^2$
goes to zero at the horizon \cite{Wald}.  From eqns.\ (\ref{A4}) 
and (\ref{A7}), we see that
\beq
{\rho^a\rho^b\over\rho^2}\nabla_a\rho_b = \left(
g^{ab} - {\chi^a\chi^b\over\chi^2} - \sigma^{ab} \right)\nabla_a\rho_b
= -\kappa{\rho^2\over\chi^2} + O(\chi^2) ,
\label{A8}
\eeq
where 
\beq
\sigma^{ab} 
  = g^{ab} - {\chi^a\chi^b\over\chi^2} - {\rho^a\rho^b\over\rho^2},
\label{A9}
\eeq
and I have assumed that ``spatial'' derivatives of $\rho^a$ and
$\chi^a$, that is, derivatives projected by $\sigma$, are 
$O(\chi^2)$ near the horizon.  Further, since $(\nabla_a\chi_b)
(\nabla^a\chi^b) = -2\kappa^2$ and $\omega_{abc}\omega^{abc}=0$ on 
the horizon \cite{Wald}, it follows from (\ref{A6})
that
\beq
{\rho^2\over\chi^2} = -1 + O(\chi^2) .
\label{A10}
\eeq

\section{The Hamiltonian for General Relativity \label{appenb}}

We know from eqn.\ (\ref{a8}) that the Hamiltonian for general 
relativity can be written as a sum of two terms.  The first of
these terms, $\int{\bf Q}$, was evaluated in section \ref{secc}.
In this appendix, I discuss the second term, which must be 
determined by solving eqn.\ (\ref{a9}) for the $(n-1)$-form
$\bf B$.

As in section \ref{secc}, we shall treat $\chi^a$ and $\rho_a$ as 
fixed vectors, and require that variations satisfy the boundary
conditions (\ref{c3}).  This means that $\delta\chi^2 = 0$ and
$\chi^a\delta\chi_a=0$, up to terms of order $\chi^2$ that will
drop out at the horizon.  In analogy with the boundary condition 
(\ref{c6}), let us also set the $\rho^a$ derivatives of our variations 
to zero at the boundary:
\beq
\rho^a\nabla_a(g_{bc}\delta g^{bc}) = 0 , \qquad
\rho^a\nabla_a\left( {\rho^b\delta\chi_b\over\chi^2} \right) =
\rho^a\nabla_a\left( {\delta \rho^2\over\rho^2} \right) = 0 \qquad
\hbox{at $\chi^2=0$} .
\label{B1}
\eeq
As discussed in section \ref{secc}, for the Kerr metric in 
Boyer-Lindquist coordinates this is the requirement that radial
derivatives not blow up at the horizon.

Now let $\xi^a$ be a vector of the form (\ref{c4}).  From eqn.\ 
(\ref{a11}), it is not hard to show that
\beq
16\pi G\xi^b\Th_{ba_1\dots a_{n-2}} = -{\hat\epsilon}_{a_1\dots a_{n-2}}
   (T A + R B) + O(\chi^2) ,
\label{B2}
\eeq
with 
\begin{eqnarray}
A &=& {|\chi|\over\rho}\left( \rho^c\nabla_c(g_{ab}\delta g^{ab})
   - \rho_b\nabla_c\delta g^{bc} \right)
   = {|\chi|\over\rho}\left(\nabla_b\rho_c\delta g^{bc} 
   - \nabla_a\delta\rho^a\right) \nonumber \\
B &=& {\rho\over|\chi|}\left( \chi^c\nabla_c(g_{ab}\delta g^{ab})
   - \chi_b\nabla_c\delta g^{bc} \right)
   = {\rho\over|\chi|}\left( \chi^c\nabla_c(g_{ab}\delta g^{ab})
   + g^{bc}\nabla_b\delta\chi_c \right) ,
\label{B3}
\end{eqnarray}
where I have used an argument parallel that following eqn.\ 
(\ref{c9}) to eliminate some terms involving variations tangent to
the horizon.  Using identities from \ref{appena}, one finds that
\begin{eqnarray}
A &=& {|\chi|\over\rho}\left\{ \kappa{\delta\rho^2\over\chi^2}
   + \chi^a\nabla_a\left( {\rho^b\delta\chi_b\over\chi^2} \right) 
   \right\} + O(\chi^2) \nonumber\\
B &=& {\rho\over|\chi|}\left\{ -2\delta(\nabla_a\chi^a) 
   - 2\kappa{\rho^b\delta\chi_b\over\chi^2} \right\} + O(\chi^2) .
\label{B4}
\end{eqnarray}
For a diffeomorphism satisfying the boundary condition (\ref{c5}), 
eqn.\ (\ref{B2}) thus gives
\begin{eqnarray}
\xi^b\Th_{ba_1\dots a_{n-2}} = -{1\over16\pi G}
   {\hat\epsilon}_{a_1\dots a_{n-2}}{|\chi|\over\rho}
   \Biggl\{ T\,\Biggl[&& \kappa{\delta\rho^2\over\chi^2}
   + D\left( {\rho^b\delta\chi_b\over\chi^2} \right) \Biggr] 
   \\
   && - {1\over\kappa} DT \left[ -2\delta(\nabla_a\chi^a)
   - 2\kappa{\rho^b\delta\chi_b\over\chi^2} \right] \Biggr\} 
   + O(\chi^2) , \nonumber
\label{B5}
\end{eqnarray}
or equivalently
\begin{eqnarray}
\xi^b\Th_{ba_1\dots a_{n-2}} = -{1\over16\pi G}
   {\hat\epsilon}_{a_1\dots a_{n-2}} \Biggl\{
   - T\Biggl[&& 2\kappa{\delta\rho\over|\chi|}
   + {|\chi|\over\rho}D\left( {\rho^b\delta\chi_b\over\chi^2} \right)\Biggr] 
   \\
   && + {2\over\kappa}{|\chi|\over\rho} DT\, \delta(\nabla_a\chi^a)
   + 2D\left( {|\chi|\over\rho}T\,{\rho^b\delta\chi_b\over\chi^2}\right)
   \Biggr\} + O(\chi^2) . \nonumber
\label{B6}
\end{eqnarray}

Now, if $\hat\epsilon$ were fixed---that is, if we froze the induced
metric on the boundary $\cal H$---then the variations in eqn.\ (\ref{B6})
could be pulled through the prefactor $\hat\epsilon$, and one could
write the entire expression as a variation $\delta(\xi\cdot{\bf B})$.
But this is the wrong boundary condition for a black hole horizon 
\cite{Teitelboim2,Carlip6}; one should rather hold fixed the momentum 
conjugate to the horizon metric.\footnote{In the Euclidean theory,
this conjugate variable is the deficit angle at the horizon.  Note
that variations of the horizon metric have dropped out of eqn.\ 
(\ref{B6}) because so far they have appeared only in terms of order 
$\chi^2$, not because they have been set to zero.}  In general, the 
variation $\delta\hat\epsilon$ will be independent of the variations 
$\delta\chi_a$ and $\delta\rho^a$ appearing in eqn.\ (\ref{B6}).  This 
means that $\xi^b\Th_b$ will be a total variation only if (\ref{B6}) 
can be written in the form ${\hat\epsilon}\times\delta(\hbox{\em terms 
that vanish on shell})$.

For the first two terms, this is not hard.  It is straightforward to
check that
\begin{eqnarray}
&&\delta\left({\chi^a\rho^b\over\chi^2}
   (\nabla_a\chi_b + \nabla_b\chi_a)\right) 
   = D\left({\rho^b\delta\chi_b\over\chi^2}\right) \nonumber\\
&&-{1\over2}{|\chi|\over\rho}\delta\left( {(\rho^a-\chi^a)(\rho_a-\chi_a)
   \over\chi^2}\right) ={\delta\rho\over|\chi|} ,
\label{B7}
\end{eqnarray}
and the left-hand side of both of these expressions vanishes at $\cal H$
when the boundary is a Killing horizon.  Similarly, $\nabla_a\chi^a=0$
when $\chi^a$ is a Killing vector.  We can thus write
\beq
\xi^b\Th_{ba_1\dots a_{n-2}} 
   = \delta\left(\xi^b{\bf B}_{ba_1\dots a_{n-2}}\right)
   -{1\over8\pi G}{\hat\epsilon}_{a_1\dots a_{n-2}} D\Biggl(
   {|\chi|\over\rho}T\,{\rho^b\delta\chi_b\over\chi^2}\Biggr) 
   + O(\chi^2)
\label{B8}
\eeq
with
\begin{eqnarray}
\xi^b{\bf B}_{ba_1\dots a_{n-2}} = - {1\over16\pi G} 
   {\hat\epsilon}_{a_1\dots a_{n-2}}\Biggl\{ {|\chi|\over\rho}T\,
   \Biggl[ \kappa&&{(\rho^a-\chi^a)(\rho_a-\chi_a)\over\chi^2} \\
   &&\ - {\chi^a\rho^b\over\chi^2}(\nabla_a\chi_b + \nabla_b\chi_a)\Biggr] 
   + {2\over\kappa}{|\chi|\over\rho}DT\,\nabla_a\chi^a \Biggr\} . 
   \nonumber
\label{B9}
\end{eqnarray}
   
It remains for us to deal with the last term in eqn.\ (\ref{B8}).  In
general, this expression cannot be written as a total variation unless 
we either strengthen the boundary conditions or further restrict the 
allowed variations of the metric.  The basic problem is that the
quantity $\rho^b\delta\chi_b/\chi^2$ is not itself the variation of 
a local function.  Indeed, the commutator
\beq
\delta_2\left( {\rho^b\delta_1\chi_b\over\chi^2} \right) -
\delta_1\left( {\rho^b\delta_2\chi_b\over\chi^2} \right)
= {\delta_2\rho^2\over\rho^2}{\rho^b\delta_1\chi_b\over\chi^2} 
- {\delta_1\rho^2\over\rho^2}{\rho^b\delta_2\chi_b\over\chi^2} 
+ O(\chi^2)
\label{B10}
\eeq
would have to vanish if $\rho^b\delta\chi_b/\chi^2$ were a total
variation.  But it is evident that this quantity is not in general 
zero, since $\delta\rho^2$ and $\rho^a\delta\chi_a$ can be
specified independently.  

We now have three choices.  First, we can try to strengthen our 
boundary conditions, for instance by fixing $\chi_a$ at $\cal H$, 
to allow the Hamiltonian to be defined.  Fixing $\chi_a$ is too 
strong a restriction, though---it is incompatible with the existence 
of diffeomorphisms of the form (\ref{c5})---and it seems difficult 
to find an alternative weak enough to allow any interesting central 
extensions of $\hbox{Diff}M$ to remain.  Second, we can consider 
``integrated generators'' $\int\! dv\,H[\xi]$, as introduced by 
Cadoni and Mignemi \cite{Cadoni} and discussed briefly in section 
\ref{secd}.  The $v$ integral would then eliminate the last term 
in eqn.\ (\ref{B8}).  But as noted in section \ref{secd}, the meaning
of such generators is unclear in the present context. 

Our third alternatively is to restrict our field variations to 
bring the last term in eqn.\ (\ref{B8}) under control.  To analyze
this possibility, let us consider mode expansions of $T$ and 
$\rho^b\delta\chi_b/\chi^2$,
\beq
T = \sum_n T_ne^{i\kappa n v} , \qquad
{\rho^b\delta\chi_b\over\chi^2} = \sum_n b_n e^{i\kappa n v} ,
\label{B11}
\eeq
where the period $2\pi/\kappa$ has been chosen for convenience.
The term in question then consists of a sum of pieces of the form
$$ (m+n)\int_{\cal H}{\hat\epsilon}b_m T_n e^{i\kappa(m+n)v} ,$$
and will vanish if
\beq
\int_{\cal H}{\hat\epsilon}_{a_1\dots a_{n-2}} b_m T_n 
   \sim \delta_{m+n} .
\label{B12}
\eeq
With such a choice, the last term in eqn.\ (\ref{B8}) is zero, and
(\ref{B9}) gives the full $(n-2)$-form $\bf B$ needed for the
Hamiltonian in eqn.\ (\ref{a8}).

The orthogonality relation (\ref{B12}) arose from demanding the
existence of $H[\xi]$.  It would clearly be preferable to have it 
come directly from a boundary condition.  One possible condition
is that of eqn.\ (\ref{c7b}), which essentially requires that the
average surface gravity remain fixed.  To see that this boundary
condition implies (\ref{B12}), first note that
\beq
\delta\int_{\cal H}{\hat\epsilon}
   \left({\tilde\kappa} - {\rho\over|\chi|}\kappa\right)
   = -\int_{\cal H}{\hat\epsilon}D
   \left({\rho^b\delta\chi_b\over\chi^2}\right)
   = \int_{\cal H}{\hat\epsilon}D
   \left({\chi_b\delta\rho^b\over\chi^2}\right) ,
\label{B12a}
\eeq
as can easily be seen from the identities in \ref{appena}.  We
must now consider what variations $\chi_b\delta\rho^b/\chi^2$ are 
allowed.  We must certainly permit variations $\delta_\xi\rho^a 
= (\delta_\xi g^{ab})\rho_b$ corresponding to diffeomorphisms generated
by vector fields satisfying the conditions (\ref{c4}), (\ref{c5}),
and (\ref{c6}).  But for consistency, we must then also allow variations 
of the form $\delta(\delta_\xi\rho^a)$ whenever $\delta\rho^a$ is 
itself allowed.  For such variations, eqn.\ (\ref{B12a}) becomes
\beq
\int_{\cal H}{\hat\epsilon} D\left( 
   DT_1 {\rho^b\delta_2\chi_b\over\chi^2} \right) = 0 .
\label{B12c}
\eeq
Together with the mode expansion (\ref{B11}), eqns.\ (\ref{B12a}) 
and (\ref{B12c}) give (\ref{B12}), as required.  

Finally, let us verify that (\ref{B9}) gives the appropriate
contribution to the Dirac bracket (\ref{c13}).  To check this, we must
look at the variation of $\bf B$ under a second diffeomorphism that
satisfies the conditions (\ref{c4}), (\ref{c5}), and (\ref{c6}).
Under such a variation,
\begin{eqnarray}
&&g_{ab}\delta_{\xi_2}g^{ab} = -2\nabla_a\xi_2^a = 2DT_2 \nonumber\\
&&\delta_{\xi_2}\rho^a = (\delta_{\xi_2}g^{ab})\rho_b
   = -{1\over\kappa}D^2T_2\chi^a + 2DT_2\rho^a + O(\chi^2) ,
\label{B13}
\end{eqnarray}
and hence
\begin{eqnarray}
&&\delta_{\xi_2}(\nabla_a\chi^a) = -{1\over2}D(g_{ab}\delta_{\xi_2}g^{ab})
  = -D^2T_2 \nonumber\\
&&\delta_{\xi_2}\left({\chi^a\rho^b\over\chi^2}
  (\nabla_a\chi_b + \nabla_b\chi_a)\right) 
  = -D\left( {\chi_b\delta_{\xi_2}\rho^b\over\chi^2} \right)
  = {1\over\kappa}D^3T_2 + O(\chi^2) \nonumber\\
&&\delta_{\xi_2}\left( {(\rho^a-\chi^a)(\rho_a-\chi_a)
   \over\chi^2}\right) = {\delta_{\xi_2}\rho^2\over\chi|^2}
  = 2{\rho^2\over\chi^2}DT_2 + O(\chi^2) .
\label{B14}
\end{eqnarray}
Note that the vector $\chi^a$ and the one-form $\rho_a$ have been held
fixed in these variations, since they are being treated as fixed,
field-independent parameters, while $\delta_{\xi_2}$ means the 
variation induced by the Poisson brackets on the phase space.
Substituting eqn.\ (\ref{B14}) into eqn.\ (\ref{B9}), we see that
\begin{eqnarray}
\delta_{\xi_2}\int_{\cal H}&&\xi_1^b{\bf B}_{ba_1\dots a_{n-2}}\\
   &&= -{1\over16\pi G}
   \int_{\cal H}{\hat\epsilon}_{a_1\dots a_{n-2}} \left\{
   - {2\over\kappa}D\left({|\chi|\over\rho}T_1D^2T_2\right)
   +  {1\over\kappa}T_1D^3T_2 - 2\kappa T_1DT_2 \right\} + O(\chi^2) 
   \nonumber.
\label{B15}
\end{eqnarray}
By the orthogonality conditions discussed above, the first term gives no
contribution, and we find exact agreement with the terms proportional to 
$T_1$ in eqn.\ (\ref{c13}).

\section{Does the Cardy Formula Apply? \label{appenc}}

Begin with a conformal field theory on the plane.  Such a theory is
characterized by a pair of Virasoro algebras, one for left-moving
modes and one for right-moving modes, and states will fall into
representations of these algebras.  Conversely, any theory whose
states provide a representation of a Virasoro algebra has a
conformal field theoretic description.  

Since the plane is conformal to the cylinder, we can transform our 
theory to one on a cylinder; the central termis a conformal anomaly, 
but its effect on such a transformation is simply to shift the 
stress-energy tensor \cite{CFT}.  To count states, we can now use 
a standard trick: we first compute the partition function, and then 
obtain the density of states from a Legendre transformation.  We 
therefore continue our theory to imaginary time and compactify the 
cylinder to a torus of modulus $\tau$.  The partition function is then
\beq
Z(\tau,{\bar\tau}) = \hbox{Tr}\, e^{2\pi i\tau L_0}
   e^{2\pi i{\bar\tau}{\bar L}_0} = \sum \rho(\Delta,{\bar\Delta})
   e^{2\pi i\tau\Delta}e^{2\pi i{\bar\tau}{\bar\Delta}} ,
\label{C1}
\eeq
and if we can determine $Z$, we can extract the density of states
$\rho(\Delta,{\bar\Delta})$ by means of a contour integral.  It should 
be stressed that the transformation from the plane to the cylinder and 
the continuation to imaginary time are merely tricks to obtain the density 
of states; we are {\em not\/} assuming any fundamental role for compact 
spaces or Euclidean signature.

The derivation of the Cardy formula starts with the observation 
\cite{Cardy} that the quantity
\beq
Z_0(\tau,\bar\tau) = \hbox{Tr}\, e^{2\pi i\tau(L_0 - {c\over24})}
   e^{2\pi i{\bar\tau}({\bar L}_0 - {c\over24})}
\label{C2}
\eeq
is invariant under modular transformations, the large diffeomorphisms
of the torus.  In particular, $Z_0$ is invariant under the $S$
transformation $\tau\rightarrow-1/\tau$.  Using this invariance, one 
can write $Z(\tau,{\bar\tau})$ in terms of $Z(-1/\tau,-1/{\bar\tau})$
and a rapidly varying phase, and use the method of steepest descents
to extract $\rho(\Delta,{\bar\Delta})$.  Details are given in ref.\
\cite{Carlip2}; the general result is that
\beq
\rho(\Delta) \sim \exp\left\{ 
   2\pi\sqrt{{c_{\hbox{\scriptsize eff}}\over6}
   \left(\Delta - {c\over24}\right)} \right\} \rho(\Delta_0) ,
\label{C3}
\eeq
where the ``effective central charge'' is
\beq
c_{\hbox{\scriptsize eff}} = c - 24\Delta_0 .
\label{C4}
\eeq
and $\Delta_0$ is the lowest eigenvalue of $L_0$ in the trace 
(\ref{C1}).

To determine the applicability of this formula to the problem
discussed in this paper, we must check several points.  First,
the conformal field theories for which the Cardy formula was 
developed are two-dimensional and have two Virasoro algebras, 
while we have no obviously important two-manifold and have only 
one Virasoro algebra.  The derivation of the Cardy formula
(\ref{C3}), however, requires few of the details of conformal
field theory; all that is really needed is the existence of
a Virasoro algebra and the diffeomorphism invariance expressed
by eqn.\ (\ref{C2}).  In essence, one may forget about the original
physical motivation, and view the Cardy formula as a statement
about representations of $\hbox{Diff}\,S^1$.  In particular, left- 
and right-moving states in a conformal field theory effectively
decouple, and the central extension described in this paper simply 
corresponds to a conformal field  theory in which one sector is 
absent.

Second, the derivation described in this appendix implicitly assumed 
that $L_0$ had a discrete spectrum.  While much of section \ref{secd} 
was also based on a discrete set of modes---see, for example, eqn.\ 
(\ref{d3})---it is not clear that this is an appropriate assumption, 
and one might well want to recast the argument to describe a 
continuous set of modes.  This presents no difficulty, however: 
the derivation of the Cardy formula requires only small modifications 
when $L_0$ has a continuous spectrum.  In that case, one should 
understand $\rho(\Delta)d\Delta$ as a density of states in an
interval $d\Delta$ of eigenvalues of $L_0$, but the interpretation
of $S$ in eqn.\ (\ref{d13}) as entropy remains unchanged.

Finally, we must worry about the value of $\Delta_0$, and 
the difference between the central charge $c$ and the effective 
central charge $c_{\hbox{\scriptsize eff}}$ that appears in the 
Cardy formula.  Here, the methods of this paper have nothing to 
say: one can determine $\Delta_0$ and $c_{\hbox{\scriptsize eff}}$
only when one has a concrete conformal field theory to represent
the horizon degrees of freedom.

\end{document}